\documentclass[12pt]{iopart}

    \usepackage{textcomp}
	\usepackage{amsmath,amssymb}
	\usepackage{makeidx}
	\usepackage{amsfonts}
	\usepackage[ansinew]{inputenc}
	\usepackage[usenames,dvipsnames]{pstricks}
	\usepackage{subfigure}
	\usepackage{epsfig}
	\usepackage{pst-grad} 

\newcounter{concno}
\newcommand{\conc}[1]{\refstepcounter{concno} \Alph{concno}\label{#1}}

\newcommand{\be}{\begin{equation}}
\newcommand{\ee}{\end{equation}}

\newcommand\bea{\begin{eqnarray}}
\newcommand\eea{\end{eqnarray}}
\newcommand\ba{\begin{array}}
\newcommand\ea{\end{array}}

\newcommand{\refer}[1]{(\ref{#1})}
\newcommand{\dd}{\mbox{d}}

\begin{document}
\title{Existence and stability of circular orbits in general static and spherically symmetric spacetimes}

\author{Junji Jia$^{1,2}$, Jiawei Liu$^{1,2}$, Xionghui Liu$^{1,2}$, Zhongyou Mo$^{1,2}$, Xiankai Pang$^{1,2}$, Yaoguang Wang$^{2,3}$, Nan Yang$^4$}

\address{$^1$ MOE Key Laboratory of Artificial Micro- and Nano-structures, Wuhan University, Wuhan, 430072, China}
\address{$^2$ School of Physics and Technology, Wuhan University, Wuhan, 430072, China}
\address{$^3$ Institute of High Energy Physics, Chinese Academy of Sciences, Beijing 100049, China}
\address{$^4$ Glyn O. Phillips Hydrocolloid Research Centre, Hubei University of Technology, Wuhan 430068, China}

\ead{junjijia@whu.edu.cn}

\begin{abstract}
The existence and stability of circular orbits (CO) in static and spherically symmetric (SSS) spacetime are important because of their practical and potential usefulness. In this paper, using the fixed point method, we first prove a necessary and sufficient condition on the metric function for the existence of timelike COs in SSS spacetimes. After analyzing the asymptotic behavior of the metric, we then show that asymptotic flat SSS spacetime that corresponds to a negative Newtonian potential at large $r$ will always allow the existence of CO. The stability of the CO in a general SSS spacetime is then studied using the Lyapunov exponent method. Two sufficient conditions on the (in)stability of the COs are obtained. For null geodesics, a sufficient condition on the metric function for the (in)stability of null CO is also obtained. We then illustrate one powerful application of these results by showing that an SU(2) Yang-Mills-Einstein SSS spacetime whose metric function is not known, will allow the existence of timelike COs. We also used our results to assert the existence and (in)stabilities of a number of known SSS metrics.\end{abstract}


\submitto{\CQG}

\maketitle

\section{Introduction}

Static and spherically symmetric (SSS) spacetimes are the best studied spacetimes in the general theory of relativity. Birkhoff's theorem guarantees that any spherically symmetric solution of the vacuum field equations is static and asymptotically flat and the exterior solution is always given by the Schwarzschild metric \cite{Jebsen, birkhoff, hawking}. The generalization of this theorem to include charge states that any spherically symmetric and electrically charged solution is stationary and asymptotically flat and can be cast into the Reissner-Nordstr\"{o}m (RN) metric form. Moreover, enormous amount of SSS spacetimes with different energy-momentum distributions, representing real physical system to different extents have been studied in the literature (see Ref. \cite{Delgaty:1998uy,exactsolutionsofes} for a list of these solutions).

On the other hand, the study of geodesics of timelike or null test objects in these spacetimes is also very useful and interesting. Verification of general relativity in its early days depended exclusively on the using of timelike or null geodesics, including precession of the perihelion of Mercury and bending of light ray.
In studying the geodesics of the SSS spacetimes, we are often concerned with the geodesics that might be of use that is more important or practical. Among these are the circular orbits (COs) in SSS spacetimes, for a few reasons. First of all, if the CO is stable, they will permit the possibility for test objects such as satellites or spacecrafts to fall freely on such orbits with constant radius but without burning extra fuel or worrying about falling into or away from the central object. Due to this advantage, it is very desirable to know whether a given SSS spacetime permits any CO, and then adjust one's own orbit to such COs if they do exist. Secondly, COs of some special spacetime might be of great importance for astrophysics and theoretical research of gravity. For example, the inner-most (or marginally) stable COs play an important role in the accretion disk theory \cite{Abramowicz:2010nk, Chakraborty:2013kza, Ono:2015jqa} and consequently closely related to the chaotic motion \cite{Motter:2003jm} and behavior of gravitational waves originated from the central source such as black hole binaries \cite{Cornish:2003ig, Blanchet:2013haa}. Study of the COs in black hole spacetimes with electric charge and/or scalar field can also be used to test properties of these black holes such as their no-hair theorem and extremality \cite{Bambi:2008jg,Pradhan:2010ws,Johannsen:2010xs,Johannsen:2010ru,Loeb:2013lfa} and even the validity of some modified gravity \cite{Yunes:2009hc,Barausse:2011pu,Konoplya:2011qq,Yagi:2012ya}.

A CO for an SSS spacetime is mathematically equivalent to a fixed point (FP) of the radical geodesics equation.  The radical geodesic equation for the SSS metric, as a second order differential equation, can always be casted into a first order differential equation system. The FP of this system always contain one equation $\dd r/\dd \tau=0$ where $r$ is the radius of the SSS metric and $\tau$ can be chosen as the proper time or the azimuth angular coordinate. It is clear then a FP always implies a CO. In the opposite way, for a CO, we always have $\dd r(\tau)/\dd \tau=0$, $\dd^2 r(\tau)/\dd \tau^2=0$ and therefore a FP is also guaranteed. Therefore in this paper we will use the term CO and FP indiscriminately. The study of FPs, as well as the Lyapunov method that will be used to study their stabilities, are mathematically part of the {phase space analysis} theory. In the literature, there are a few works \cite{Cornish:2003ig,Palit:2008vb, Cardoso:2008bp, Pradhan:2012bh, Dean:2013jza} that applied these methods to study COs, with their emphasis on various different properties or applications of these orbits. However all these works are done to spacetimes with explicit metric functions, such as Schwarzschild(-de Sitter), RN and some axially symmetric spacetimes.

In this paper, we study the condition on the general SSS spacetime for the (non-)existence of FPs for both timelike and null geodesics, and then the stabilities of these FPs.
The paper is organized in the following way. In section \ref{metgeo} we build up the metric and geodesic equations and the FP equation system. In section \ref{secexistence} we give one of the main result on the (non-)existence of the FP for SSS metrics. Then in section \ref{secexflat} we put more stringent conditions on the SSS metric and derive the conclusion that an asymptotically flat SSS metric that has a negative corresponding Newtonian potential at infinite radius will always admit FPs. The stability of the FPs for SSS spacetimes then is discussed in section \ref{secstb}. The existence and stability of the FPs for null geodesics then are studied in section \ref{secnull}. Application of these results and possible extensions are finally discussed in \ref{secdis}.

We emphasis that even though the COs of some important SSS spacetime with known metric functions are well studied,  unlike any previous analysis, in our work on the existence and stability of these COs we do not need to know the metric function {\it a priori}. Rather, the criterions we established for the (non-)existence of FPs can even be applied to cases that the metric functions are not completely known. Therefore, our results given in the form of a few theorems are not only new but vastly applicable to a large number of metrics. The power of these results is illustrated using the examples in the Discussion section.

\section{Metric and geodesic equations \label{metgeo}}

In this section we set up our notation for the metric and derive the basic geodesic equations associated with it. We also derive explicitly the FP autonomous equation system from the radial geodesic equation. The most general form of the SSS spacetime metric can be chosen as the following
\be \dd s^2=f(r)\dd t^2-\frac{1}{g(r)}\dd r^2-r^2\dd \Omega^2_{d-2}, \label{metric}\ee
where $(t,~r,~\theta,~\phi)$ are the coordinates, $\dd \Omega^2_{d-2}$ is the solid angle element of the $d-2$ dimensional sphere where $d$ is the total dimension of the spacetime.
Using this metric, the geodesic equations can be routinely derived. Due to the static and spherical symmetry of the metric, we can always set in the geodesic equations $\theta(\tau)=\pi/2$, where we use $\tau$ to denote proper time for timelike geodesics and the affine parameter for null geodesics. Moreover, there always exist two first integrals whose constants are identified with the energy $E$ and angular momentum $L$ of the test object per unit mass
\be \frac{\dd\phi(\tau)}{\dd \tau}=\frac{L}{r^2},~\frac{\dd t(\tau)}{\dd \tau}=\frac{E}{f(r)}. \label{dtdtaueq}\ee
The equation for $r(\tau)$ then is
\be \left[\frac{\dd r(\tau)}{\dd \tau}\right]^2=g(r)\left[\frac{E^2}{f(r)}-\frac{L^2}{r^2}-\delta_n\right]\equiv V(r),\label{rgeo}\ee
where $\delta_n=1,~0$ for timelike and null geodesics respectively, and $V(r)$ is defined as the entire right hand side of Eq. \refer{rgeo}.
Taking derivative of Eq. \refer{rgeo} with respect to $\tau$ again, we obtain
\be \frac{\dd^2 r(\tau)}{\dd \tau^2}= \frac{V^\prime(r)}{2},\label{rgeodiff}\ee
where $^\prime$ here and henthforce denotes the derivative with respect to $r$.

Eqs. \refer{rgeo} and \refer{rgeodiff} combined can be thought as an autonomous system of two first order differential equations for $x(\tau)=r(\tau)$ and $y(\tau)\equiv\dd r(\tau)/\dd \tau$.
Now the FP for this system is at a radius, denoted by $x=x_*$, that satisfying
\bea && \frac{\dd x(\tau)}{\dd \tau}=y(\tau)=0, \label{autoeq1}\\
 && \frac{\dd y(\tau)}{\dd \tau}=\frac{V^\prime(x_*)}{2}=0 \label{autoeq2}\eea
at some instantaneous time and all times after.
Noting \refer{rgeo}, this is equivalent to require that at $x_*=r_*$
\be V(r_*)=0,~~~\mbox{and}~~~ V^\prime(r_*)=0. \label{condv}\ee
For the timelike geodesics, using \refer{rgeo} again these become
\be
\left(
\ba{cc}
r_*^2     &-f(r_*)\\
r_*^3f^\prime(r_*) & -2f(r_*)^2 \\
\ea
\right)
\left(
\ba{c}
E^2\\
L^2
\ea
\right)
=
\left(
\ba{l}
r_*^2f(r_*)\\
0
\ea
\right)
.\label{eqmat}\ee
Treating $E^2$ and $L^2$ as unknowns, this is a simple linear system. It is easy to check that when and only when
\be 2f(r_*)-r_*f^\prime (r_*)=0 \label{nosolrank}\ee
the rank of the augmented matrix is larger than that of the coefficient matrix and then there exist no solution. Otherwise, these equations can be transformed to
\bea
E^2&=& \frac{2 f^2(r_*)}{2f(r_*)-r_*f^\prime (r_*)}, \label{eqone}\\
L^2&=& \frac{r_*^3 f^\prime(r_*)}{2f(r_*)-r_*f^\prime (r_*)}. \label{eqtwo}
\eea
These equations are in agreement with Ref. \cite{Cardoso:2008bp}. The corresponding equations in the null geodesic case will be consider in section \ref{secnull}.

In this paper, we assume that the metric function $f(r)$ is positive in the range of $r$ in which we seek the FP, because otherwise according to the metric \refer{metric} the $r$ coordinate should have to be interpreted as time, not radius and a FP in time is not of our interest. We also assume that in this range of $r$ the spacetime is not singular because otherwise we cannot study the existence of FP on singular radius anyways. And then we know that there always exist coordinate transforms to make $f(r)$ continuous if it was not in the first place. Therefore throughout this paper we assume that $f(r)$ is already made continuous.

\section{Existence of the Fixed Points\label{secexistence}}

For a given $f(r)$, if in the space spanned by energy $E$ and angular momentum $L$ there exist a non-empty set $S$ for whose element $(E,~L)$ the solution to Eq. \refer{eqmat} does exist at some radius $r=r_*$, then we say for that $f(r)$ the FPs can exist (for that $(E,~L)$ at $r=r_*$). If for a given $f(r)$, the set $S$ is empty, then we say that there exist no FP for the spacetime described by $f(r)$. Apparently, if FPs exist for some $(E,~L)$ then the two equations in Eq. \refer{eqmat} will both have solutions simultaneously. If either of \refer{eqone} or \refer{eqtwo} is not satisfied by any r, then the FPs do not exist for that $(E,~L)$. Under these definitions, it is clear then if $f(0)>0$, then there always exists a timelike FP at $r_*=0, ~E=\sqrt{f(0)},~L=0$. This FP in most physically important cases is not relavent, either due to the singularity of the $r=0$ point, or due to the presence of matter at $r=0$ which prevents the test particle from doing geodesic motion, or it is just a trivial FP such as in the case of Minkovski spacetime. Therefore in the following sections when the FPs of timelike geodesics are discussed, FP at this point is excluded and we concentrate on non-trivial ones. Now let us study Eq. \refer{eqone} and \refer{eqtwo} separately first and then combine the results to obtain our main conclusion presented as Theorem \ref{nopiece}.

Keeping in mind that $E$ is real, it is clear that as long as the denominator $2f -rf^\prime$ of Eq. \refer{eqone} is positive for the given $f(r)$ and some $r$, then there always exist $E^2$ such that Eq. \refer{eqone} is satisfied. The contrapositive statement is that, no $E^2$ will satisfy Eq. \refer{eqone} if and only if for all $r$ the denominator  $2f -rf^\prime$  will remain negative. In this case, and noticing another case \refer{nosolrank}, we can assume that for all $r$,
\be 2f -rf^\prime=-\delta(r) \label{eq1fund} \ee
where $\delta(r)$ is a semi-positive but otherwise arbitrary function. Due to its arbitrariness and for the sake of a shorter expression for the solution to \refer{eq1fund} , we can also change the function $\delta(r)$ to $r^3\dd \kappa(r)/\dd r$ without losing any generality as long as $\kappa(r)$ is monotonically increasing. Then the Eq. \refer{eq1fund} becomes
\be \frac{\dd f(r)}{\dd r}-2\frac{f}{r}=r^2\frac{\dd \kappa(r)}{\dd r} \label{eq1fund2} \ee
whose solution is readily found to be
\be f(r)=(\kappa(r)+c_\kappa)r^2, \label{eq1fsol1pre}\ee
where $c_\kappa$ is an arbitrary constant. Since the only requirement for $\kappa(r)$ is that it is increasing, then we can always redefine $\kappa(r)$ such that $\kappa(r)+c_\kappa\to \kappa(r)$. Finally, we get
\be f(r)=\kappa(r)r^2. \label{eq1fsol1}\ee
From this and the requirement that $f(r)>0$, we establish the statement that: if the metric function $f(r)$ is expressed as \refer{eq1fsol1} for any positive and monotonically increasing function  $\kappa(r)$, then Eq. \refer{eqone} will have no solution at any $r$. Some remarks should be emphasized here. First, it is obvious that the opposite is also true: if Eq. \refer{eqone} is not satisfied by any $r$, then $f(r)$ must be expressible as \refer{eq1fsol1}. Secondly, the form \refer{eq1fsol1} is valid also locally: in the neighborhood of any $r$ such that the denominator $2f -rf^\prime$ is negative, $f(r)$ should be expressible as \refer{eq1fsol1}, for some locally increasing $\kappa(r)$. Lastly, the contrary is also true: if the $\kappa(r)$ in \refer{eq1fsol1} is locally positive and increasing, then Eq. \refer{eqone} is not satisfied in the corresponding neighborhood.

Now Eq. \refer{eqtwo},  for a given $f(r)$, will not be satisfied by any $r$ if and only if one of the following two conditions are satisfied: (a) for the $r$ such that the denominator $2f -rf^\prime<0$, the factor in the numerator $f^\prime>0$; (b) for the $r$ such that the denominator $2f -rf^\prime>0$, $f^\prime<0$ is true. For case (a), the solution to $2f -rf^\prime<0$ is still formally given by \refer{eq1fsol1} but only in the neighborhood of $r$ such that $f^\prime>0$. Using \refer{eq1fsol1}, one finds
\be f^\prime =\kappa^\prime(r) r^2+2\kappa(r)r.\ee
Since $\kappa(r)$ is an increasing function and positive, we see that $f^\prime>0$ is automatically satisfied. For case (b), since $f^\prime<0$ will automatically guarantee that $2f -rf^\prime>0$ because $f>0$, the only requirement therefore is
\be f^\prime<0 \label{eq1fsol2}.\ee

Combining the above two cases for Eqs. \refer{eqone} and \refer{eqtwo}, we know that the FPs of the metric will not exist if and only if either in the entire range of $r$,  Eq. \refer{eq1fsol1} is satisfied so that Eq. \refer{eqone} is broken, or in some range of $r$ Eq. \refer{eq1fsol1} is satisfied and in the rest of the range Eq. \refer{eq1fsol2} is satisfied so that Eq. \refer{eqtwo} is broken. In the following we further prove a theorem that will remove the possibility that Eq. \refer{eq1fsol1} and \refer{eq1fsol2} are piece-wisely satisfied: \\
(Theorem \conc{nopiece}) The FPs for the given metric \refer{metric} do not exist if and only if either only Eq. \refer{eq1fsol1} is satisfied in the entire range of $r$, or only Eq. \refer{eq1fsol2} is satisfied in the entire range of $r$, but not because Eq. \refer{eq1fsol1} and \refer{eq1fsol2} are piece-wisely satisfied. \\
The proof is indeed simple. Let us assume that Eq. \refer{eq1fsol1} is satisfied in some range of $r\in(a,b)$ and Eq. \refer{eq1fsol2} is satisfied in the range of $(b, c)$. Then in $(a,b)$ and $(b,c)$ we have respectively $f^\prime>0$ and $f^\prime<0$, and consequently at $r=b$, $f^\prime(b)=0$. Now let us show that the above assumptions are impossible by considering how $f(r)$ changes in a small neighborhood $r\in(b-\chi,b]$. In this neighborhood according to \refer{eq1fsol1}, the function $f(r)/r^2=\kappa(r)$ should have a semi-positive derivative. Evaluate the following derivative at the point $r=b-\epsilon\in (b-\chi,b]$
\bea
\kappa^\prime(r)\vert_{r=b-\epsilon}&=&\left(\frac{f(r)}{r^2}\right)^\prime\bigg\rvert_{r=b-\epsilon}\nonumber\\
&=&\left[\frac{f^\prime}{r^2}-\frac{2f}{r^3}\right]\bigg\rvert_{r=b-\epsilon}\label{etad}\\
&=&\frac{f^\prime(b)}{b^2}+\left[\frac{f^{\prime\prime}(b)}{b^2} -\frac{2f^\prime(b)}{b^3}\right](-\epsilon)-\frac{2f(b)}{b^3}\nonumber\\
&=&+\left[-\frac{2f^\prime(b)}{b^3}+\frac{6f(b)}{b^4}\right](-\epsilon)+{\cal O}(\epsilon^2)\label{taylorexp}\\
&=&-\frac{2f(b)}{b^3}-\left[\frac{f^{\prime\prime}(b)}{b^2}+\frac{6f(b)}{b^4}\right]\epsilon +{\cal O}(\epsilon^2)\label{eq0sub}\\
&<&0,
\eea
where from \refer{etad} to \refer{taylorexp} the Taylor expansion to the first order of $\epsilon$ is used and in \refer{eq0sub} we substituted $f^\prime(b)=0$. Clearly to the leading (zeroth) order of $\epsilon$, $ \kappa^\prime(r)\rvert_{r=b-\epsilon}<0$, which conflicts with the requirement that $\kappa^\prime(r)\geq 0$.
Similarly to this case, if Eq. \refer{eq1fsol1} and \refer{eq1fsol2} are satisfied respectively in $r\in(b,c)$ and $(a, b)$, the same calculation can also be carried through except the neighborhood we will consider in this case becomes $r\in[b,b+\chi)$ and $-\epsilon\to\epsilon$. We will still have a confliction on the monotonicity of $\kappa(r)$.  Therefore only one of \refer{eq1fsol1} and \refer{eq1fsol2} can be  satisfied on the whole range of $r$ but not both in a piece-wise way. \\
This theorem greatly simplifies the scenarios that we need to consider when finding metrics that do not allow FPs and we will refer to it as the {\it timelike FP non-existence theorem} (TFPNET).

In the following, we give two simple examples satisfying respectively  Eq. \refer{eq1fsol1} and \refer{eq1fsol2} in the entire range of $r$ and therefore do not permit any FPs, and one more example that satisfies \refer{eq1fsol1} in some range of $r$ and \refer{eq1fsol2} in some other range and neither of them in the rest where FPs lies. The first example is given
simply by $\kappa(r)=r$ so that $f(r)=r\cdot r^2=r^3$. The resulting Eq. \refer{eqone} and \refer{eqtwo} become
\be E^2=-2r^3, ~ L^2=-3r^2. \ee
Both the two equations permit no solution and therefore no FPs. The second example is given by $f(r)=1/r$ and the Eq. \refer{eqone} and \refer{eqtwo} become
\be E^2=\frac{2}{3r},~L^2=-\frac{r^2}{3}.\ee
The second equation eliminates the existence of FPs. The third example is a simple combination of the above two:
\be f(r)=\frac{r}{1+r^4}\cdot r^2. \label{eg3}\ee
At very small and large $r$, this resembles the first and second examples respectively and therefore we expect that the FPs will not exist there. However, in the middle range, as we can expect using the TFPNET theorem,  the FPs do exists.  Substituting Eq. \refer{eg3} into \refer{eqone} and \refer{eqtwo} we obtain
\be E^2=\frac{2r^3}{3r^4-1},~L^2=-\frac{(r^4-3)r^2}{3r^4-1}.\ee
Clearly, these two equations will have solution when $r\in(3^{-1/4}, 3^{1/4})$ for some $(E,~L)$.

\section{Existence of fixed points for asymptotically flat spacetime\label{secexflat}}

Even though the previous results shows that there exist metric that has no FPs for any $(E,~L)$, the required $f(r)$ in these cases are non-trivial. Usually people are more interested in asymptotically flat spacetimes due to their physical relevance. Therefore in this section we discuss under the asymptotic flatness requirement about what will happen to the TFPNET theorem we obtained in last section.

For the result \refer{eq1fsol1} since $\kappa(r)$ is positive and increasing, apparently $f(r)$ will diverge at $r\to \infty$ and therefore do not correspond to any asymptotically flat spacetime. The case \refer{eq1fsol2} however do permit asymptotically flat spacetime. An example satisfying \refer{eq1fsol2}  would be
\be f(r)=c+\frac{2a}{r^\beta}, \label{potex}\ee
where $c>0,~a>0,~\beta>0$. It is known \cite{asybook} that a necessary and sufficient condition for a spacetime to be asymptotically flat is that its metric takes the above form with $\beta\geq1$ as $r\to\infty$. On the other hand, the Eqs. \refer{eqone} and \refer{eqtwo} corresponding to \refer{potex} become
\be E^2=\frac{r^{-\beta}(r^\beta+2a)^2}{cr^\beta+a(2+\beta)}, ~  L^2=-\frac{a\beta r^2}{cr^\beta+a(2+\beta)},\ee
which clearly exclude the existence of any FPs. Therefore, enforcing the asymptotic flatness condition alone would not make sure that all SSS spacetimes permit FPs.

On the other hand, it is well known in perturbation theory of general relativity that for an SSS spacetime, at the region of $r\to\infty$ an effective Newtonian potential $\phi(r)$ can always be assigned to the $g_{00}$ component of the metric by \cite{book:hartle}
\be 1+2\phi(r)=\lim_{r\to\infty}g_{00}(r)=\lim_{r\to\infty}f(r). \ee
The potential corresponding to example \refer{potex} would be
\be \phi(r)=\frac{a}{r^\beta}. \ee
For $a>0$, this is a positive gravity potential that can only be generated by negative mass and it would produce a repelling force to normal test objects, resulting naturally no CO. Inspired by this observation, we can further exclude metric like this by imposing that the corresponding Newtonian potential at large $r$ is negative and obtain the following theorem: \\
(Theorem \conc{flatneg}) Any static, spherically symmetric and asymptotically flat spacetime corresponding at large radius to a negative Newtonian potential would allow the existence of FP in the space of $(E,~L,~r)$.

\section{Stability of the fixed points\label{secstb}}

For asymptotically flat, SSS spacetime with negative Newtonian potential at large radius, then the above theorem \ref{flatneg} clearly assert that there will exist FP in the space spanned by $(E,~L,~r)$. For the spacetimes considered in this section, we assume that this is the case and denote the FP radius as $r_*$. Then one important question following is the stability of the FPs. Here we address this issue using the Lyapunov exponent method by following closely Ref. \cite{Cardoso:2008bp}. In the simple SSS spacetime case, the Lyapunov exponents are proportional by a positive factor to the eigenvalues of the coefficient matrix of the linear perturbation equation of the autonomous system of Eqs. \refer{autoeq1} and \refer{autoeq2}.

First of all, notice that Eqs. \refer{eqone} and \refer{eqtwo} can be considered as algebraical constraints for variables appearing in them but not differential equations of $f(r)$ as a function of $r$, simply because that these two equations are only satisfied at the FP $r_*$ but not other radius values. One sees then for any given $f(r)$, among the three inputs $\{ E, ~L,~r\}$ into these two equations, generally once any one of them is fixed, the other two will be solvable from these two constraints, if the solution does exist. For the simplicity of the argument, let us choose $E$ as the free variable, i.e., once a proper $E$ that allows the existence of the FP is chosen, these two constraints will force $L$ and $r$ to be at some particular values $L_*$ and $r_*$ respectively. Alternatively, Eqs. \refer{eqone} and \refer{eqtwo} can also be thought as relations that $f(r_*)$ and $f^{\prime}(r_*)$ satisfy in terms of $E,~L_*$ and $r_*$. In the following, we will use both these two interpretations.

For the metric \refer{metric}, the Lyapunov exponents have been computed in Ref. \cite{Cardoso:2008bp} and here we directly quote as
\be \lambda=\pm\sqrt{\frac{V^{\prime\prime}(r)}{\dd t(\tau)/\dd \tau}}. \label{lexpdef}\ee
Substituting Eqs. \refer{dtdtaueq} and \refer{rgeo} and using Eqs. \refer{eqone} and \refer{eqtwo}, one finds for the timelike case
\bea  \lambda&=&\pm\sqrt{-\frac{g}{2f}\left[\frac{3f\cdot f^\prime}{r_*}-2(f^\prime)^2+f~f^{\prime\prime}\right]} \nonumber\\
&\equiv&\pm \sqrt{g\cdot h\left(r_*, ~f, ~f^\prime, ~f^{\prime\prime}\right)/2} \eea
where all of $f,~f^\prime, ~f^{\prime\prime}$ are evaluated at $r_*$ and $h$ is defined as
\be h\left(r_*, ~f, ~f^\prime, ~f^{\prime\prime}\right)=-f^{\prime\prime}+\frac{2(f^\prime)^2}{f}-\frac{3f^\prime}{r_*}. \label{hdef} \ee
It is known that if $\lambda$ is imaginary (or real) then the FP is stable (or unstable). In order to obtain some useful conditions regarding the stability of the FP, we can study the conditions under which either
\be h(r_*)<0 \label{hlzero}\ee
or
\be h(r_*)>0.\label{hgzero} \ee
Given that $g>0$ for all $r$,  the FP will be stable (or unstable) if and only if $h(r_*)<0$ (or $h(r_*)>0$) holds true. Unfortunately, the sign of $h$ cannot be completely fixed by just using the FP constraints \refer{eqone} and \refer{eqtwo} due to the presence of $f^{\prime\prime}$ term, which does not appear in Eqs. \refer{eqone} and \refer{eqtwo} at all. Indeed, solving $f$ and $f^\prime$ from \refer{eqone} and \refer{eqtwo} and substituting into \refer{hdef}, one finds
\be h=-f^{\prime\prime} (r_*) +\frac{2(L_*^2-3r_*^2)E^2L_*^2}{(L_*^2+r_*^2)^3}, \label{hsub1} \ee
from which $f^{\prime\prime}$ cannot be removed.

Even though we cannot find the {\it necessary and sufficient} condition on the stability or instability of the FP, if we think of $h$ to be positive or negative not only for the FP radius $r=r_*$ but for all $r$, then Eqs. \refer{hlzero} or \refer{hgzero} can be treated as differential equations. This way we are able to get the following {\it sufficient} condition: \\
(Theorem \conc{stabsuff}) If the differential equation
\be -f^{\prime\prime}+\frac{2(f^\prime)^2}{f}-\frac{3f^\prime}{r}\equiv \sigma(r)<0~( \mbox{or }>0) \label{sigmadef} \ee
for all $r$, then the FP, if exists,  would be stable (or unstable).\\
In what follows we analyze only the case for $\sigma(r)<0$ and the case for $\sigma(r)>0$ will be similar. Now similar to the procedure that is used in section \ref{secexistence}, we can solve Eq. \refer{sigmadef} for $f(r)$ and its solution will always make the Lyapunov exponent imaginary and consequently the FP stable since $\sigma(r)<0$ at any radius, including $r_*$.
It is also clear that if $\sigma(r)$ is positive for all $r$, then the opposite happens: the FP will always be unstable.

Eq. \refer{sigmadef} though looks simple, has proven to be difficult to get an analytical solution. This is mainly because of the presence of $\sigma(r)$ as an arbitrary and non-homogenous term. This arbitrariness of $\sigma(r)$ however can also be taken advantage of. Without losing any generality, we can transform $\sigma(r)$ to terms proportional to the second or third term on the left hand side of Eq. \refer{sigmadef} so that the equation becomes homogenous and can be solved exactly. Each of the solved explicit form of $f(r)$ will be a sufficient condition for the (in)stability of the FP. For the first transform from $\sigma(r)$ to a term proportional to $(f^\prime)^2/f$, noticing that at the FP, $f^\prime(r_*)>0$, $f(r_*)>0$,  we can rewrite $\sigma(r)$ so that \refer{sigmadef} becomes
\be -f^{\prime\prime}+\frac{2(f^\prime)^2}{f}-\frac{3f^\prime}{r}=\frac{r^3(f^\prime)^2\zeta^\prime}{f(r)},\ee
where if $\zeta^\prime(r)<0$, the FP will be stable.
Notice that in this transform, we made sure the last term has the same sign as the $\sigma(r)$ term.
This equation can be easily solved to obtain the solution
\be f(r)=c_1\exp\left(\int\frac{2}{r[2r^2(\zeta(r)-c_2)+1]}\dd r\right).\ee
Since $\zeta(r)$ is only required to be monotonic, then the $c_2$ can be absorbed into $\zeta(r)$. We get
\be f(r)=c_1\exp\left(\int\frac{2}{r[2r^2\zeta(r)+1]}\dd r\right). \label{fsuff1}\ee
The second transform is to write \refer{sigmadef} as
\be -f^{\prime\prime}+\frac{2(f^\prime)^2}{f}-\frac{3f^\prime}{r}=- f^\prime \frac{\dd \ln\left[r^3\chi^\prime(r)\right]}{\dd r},\ee
and we require that
$r^3\chi^\prime(r)$
is a positive function with a positive derivative but otherwise arbitrary function. This will grantee that $\sigma(r)$ will be negative for all radius and therefore the FP is stable. Moreover, the equation also becomes solvable and we obtain
\be f(r)= \frac{1}{c_1\chi(r)+c_2}, \label{fsuff2}\ee
where $c_1\neq0$ and $c_2$ are constants.
If $f(r)$ can be written as \refer{fsuff1} in which $\zeta^\prime(r)<0$, or as \refer{fsuff2} in which $r^3\chi^\prime(r)$ is positive and increasing, then clearly the FPs will always be stable. On the contrary, if $\zeta(r)$ in \refer{fsuff1} is increasing or $r^3\chi^\prime(r)$ is decreasing, then the corresponding $\sigma(r)$ in \refer{sigmadef} will be positive and the FPs will be unstable.

It is worth emphasizing that the above two $f(r)$'s are just some sufficient conditions obtained using very special form of $\sigma(r)$. It is very possible that there exist other $\sigma(r)$ and consequently $f(r)$ which will make the FP stable or unstable too. Moreover, Eq. \refer{sigmadef} itself is also the consequence of a quite strong requirement that at all radius $h(r_*,~f,~f^\prime,~f^{\prime\prime})$ has a fixed sign. It is not hard to come up metrics that will only make $h$ negative (or positive) locally, i.e., near the FP but not entire $r$, which then can make the FP stable (or unstable).

\section{Fixed points of null geodesics \label{secnull}}

For null geodesics, using Eq. \refer{condv} the existence of FPs would require \cite{Cardoso:2008bp}
\bea
&&\frac{E}{L}=\pm\sqrt{\frac{f(r_*)}{r_*^2}} ,\label{eqthree}\\
&&2f(r_*)=r_*f^\prime(r_*). \label{eqfour}
\eea
An inspection of these two equations and \refer{eqone} and \refer{eqtwo} indicate that the null case is the limit of $2f(r_*)\to r_*f^\prime(r_*)$ for the timelike case.

Clearly, there always exist $(E,~L,~r)$ satisfies Eq. \refer{eqthree}. For Eq. \refer{eqfour}, the existence of FP suggest that there exist some $r_*$ that will satisfy this equation. Therefore a non-existence condition of FP would be that there exist no $r_*$ satisfying Eq. \refer{eqfour}. In other words,
\be
2f(r)-rf^\prime(r)=-\delta(r)<0 \label{eqnone} \ee
for all $r$ where $\delta(r)$ is a positive but otherwise arbitrary function or
\be
2f(r)-rf^\prime(r)=\rho(r)>0 \label{eqntwo} \ee
for all $r$ where $\rho(r)$ is an positive but otherwise arbitrary function.
Eq. \refer{eqnone} is similar to Eq. \refer{eq1fund} and its solution is
\be
f(r)=\mu(r)r^2\label{soleq1}\ee
where $\mu(r)$ is a positive function with a positive derivative. Eq. \refer{eqntwo} differs from Eq. \refer{eqnone} only by a sign and therefore the solution can be similarly solved as
\be
f(r)=\xi(r)r^2\label{soleq2}\ee
where $\xi(r)$ is a positive function with a negative derivative.
Therefore we have the following theorem:\\
(Theorem \conc{nullexistence}) The FP for null geodesics does not exist if and only if $f(r)$ takes the form of \refer{soleq1} or \refer{soleq2}.

Comparing with the TFPNET, Eq. \refer{eq1fsol2} implies that its $f(r)/r^2$, the function corresponding $\xi(r)$ in \refer{soleq2}, will always have a semi-negative derivative. This means that for the $\xi(r)$'s that have a strict negative derivative everywhere, there will exist neither timelike nor null FP. While for the $\xi(r)$ that might has zero derivative at some radius, there is still no timelike FP but there can exist null FP.
An example can be constructed as
\be \kappa(r)=a^3+(r-a)^3,~f(r)=\kappa(r)r^2=[a^3+(r-a)^3]r^2,~(a>0)\ee
which has $\kappa^\prime(r)=3(r-a)^2\geq0$ and therefore it respects \refer{eq1fsol1} and has no timelike FP. It is also clear that at $r=a$, $2f-rf^\prime=0$ and therefore there exist a null FP. On the other hand, one can also give examples of metrics that allows timelike FP but no null FPs, such as
\be
\xi(r)= \left(1+e^{-r}+\frac{1}{r^2}\right)/3,\ee
which leads to
\be f(r)=\xi(r)r^2=(1+r^2e^{-r}+r^2)/3. \ee
Here $\xi^\prime(r)<0$ and therefore respecting \refer{soleq2} while $f(r)$ violates \refer{eq1fsol2}. It is straight forward to verify that for this $f(r)$ which satisfies Eq. \refer{eqfour}, Eqs. \refer{eqthree} and \refer{eqfour} allow no physical solution for $(E,~L,~r)$ while Eq. \refer{eqone} and \refer{eqtwo} still allow real solution.

When asymptotic flatness condition is imposed on the spacetime, in general it will not guarantee the existence of FP of null geodesics. Indeed, it is not hard to verify that the metric \refer{potex} will not allow FP for null geodesics. Furthermore, when the asymptotically flat metric is required to correspond to a negative Newtonian potential at large $r$, then clearly condition \refer{soleq1} will not be satisfied. There however still exist metrics that satisfies condition \refer{soleq2}, as long as $\xi(r)$ is chosen not to decrease too fast so that when multiplied by $r^2$, the resultant $f(r)$ can still satisfies the desired asymptotic form. An example is given by
\be
\xi(r)=\frac{1}{r(r+2M)},~f(r)=\xi(r)r^2=1-\frac{2M}{r+2M}, \ee
where $M>0$ is the ADM mass. It is clear that this metric is asymptotically flat, corresponds to negative Newtonian potential at large $r$  and still satisfies Eq. \refer{soleq2}, and therefore it has no null FP. This point is clearly different from timelike case.

Now suppose we are in a spacetime allowing the existence of a FP for null geodesics. Then the stability of the FP can again be studied using the Lyapunov exponent \refer{lexpdef}, which takes the following form in this case
\be \lambda=\pm\sqrt{\frac{g(r_*)}{2r_*^2}\left[2f(r_*)-r_*^2f^{\prime\prime}(r_*)\right]} .\ee
Again, the presence of the $f^{\prime\prime}$ term implies that in the general case the stability of the FP cannot be determined definitely without actually found the FP. We can at most obtain some sufficient conditions by forcing the term under the square root to be negative (or positive) for the entire range of $r$:
\be 2f(r)-r^2f^{\prime\prime}(r)=-\mu^\prime(r), \label{nstab}\ee
where $\mu(r)$ is an arbitrary but with its derivative sign fixed.
Unlike in the case of FP of timelike geodesics, here the Eq. \refer{nstab} can be solved exactly to produce
\be f(r)=\left(\int\frac{\mu(r)}{r^4}\dd r+c\right)r^2 .\ee
When $\mu^\prime(r)>0$ (or $\mu^\prime(r)<0$) function, then the FP determined by the above $f(r)$ will be always stable (or unstable).

\section{Discussion \label{secdis}}

In this section, we first discuss two applications of our results obtained in previous sections, and then discuss possible future works.

The first application, which is also what initially motivated us from considering the FP problem for SSS spacetimes, is to use these results to study the existence of FP for a metric whose explicit formula is not completely known.
In general relativity, even for the simplest SSS case there are many models whose field equations cannot be solved analytically due to the high non-linearity  of the Einstein equations. Some of these model metrics even lack a known numerical solution. Examples of these include the famous ``colored black hole'' for the SU(2) Yang-Mills-Einstein model \cite{Bizon:1990sr}, whose solution in the entire outer region of the black hole is only known numerically.
What happens to these models' field equations is that we often only know from physical/mathematical arguments some limited features such as boundary or asymptotic behaviors of the metric functions and other fields, but not their analytical solutions. On the other hand, the existence of FP of these models might be of theoretical or practical interests. We now illustrate with the colored black hole example that our results provide an opportunity to assert the (non-)existence of FP of the metric without having to know its analytical formula.

For the SU(2) Yang-Mills-Einstein SSS model considered in Ref. \cite{Bizon:1990sr}, if we were using a metric of the form
\be \dd s^2=f(r)\dd t^2-\left(1-\frac{2m(r)}{r}\right)^{-1}\dd r^2-r^2\dd \Omega^2_2, \label{cbhmetric}\ee
the field equation for the metric function $f(r)$ becomes
\be
f^\prime(r)=4w^\prime(r)^2\frac{f(r)}{r}+2f(r)\frac{m(r)-m^\prime(r)r}{r\left[r-2m(r)\right]},\label{cbhfeq} \ee
while the field equations for the mass function $m(r)$ and gauge field $w(r)$ is still given by Eqs. (10a) and (12) of the same paper. It is also known from asymptotic analysis that for the mass and gauge functions respectively
\be m(r)\to M,~\mbox{ and }~ |w(r)|\to 1-c/r ~\mbox{ as }r\to\infty ,\label{cbhasy} \ee
where $0<M<\infty$ and $c>0$.
For these equations, the author then used the shooting method to find their numerical solution. Up to this point, the analytical solution of this metric is still not known yet.  Now in our work however we show that without knowing the solutions of Eq. \refer{cbhfeq} but with only the equation itself and the asymptotics \refer{cbhasy}, using theorems obtained previously we can prove the existence of FP for this metric.

First of all, we see that the first term on the right hand side of Eq. \refer{cbhfeq} is positive definite. For the second term, at large $r$ its sign is determined by the numerator, which we set to the following
\be m(r)-m^\prime(r)r=-p^\prime(r)r^2. \label{cbh2t}\ee
Solving this, we obtain
\be m(r)=p(r)r. \label{cbhmsol}\ee
Now due to the boundness \refer{cbhasy} of $m(r)$, clearly $p(r)\to M/r+{\cal O}(r^{-b})$ where $b>1$ as $r\to\infty$. Then we see that $p^\prime(r)\to -M/r^2+{\cal O}(r^{-b-1})<0$ and consequently from Eq. \refer{cbh2t} we see that the second term of Eq. \refer{cbhfeq} is also positive as $r\to\infty$.
Therefore $f(r)$ is an increasing function at large enough $r$ and will not satisfy Eq. \refer{eq1fsol2}. Further, we show that $f(r)$ will not increase as fast as Eq. \refer{eq1fsol1}. Substituting Eq. \refer{eq1fsol1} and \refer{cbhmsol} into \refer{cbhfeq} one finds
\be
\kappa^\prime(r)r^2+2\kappa(r)r=2\kappa(r)r\frac{2\left[1-2p(r)\right]w^\prime(r)^2-rp^\prime(r)}{1-2p(r)}.\ee
Now because of the asymptotics of $p(r)$ and $w(r)$ in \refer{cbhasy}, one see that the fraction part of the right hand side is of order $1/r$ at large $r$ and there exist no $\kappa(r)$ that can satisfy the above equation to the leading order for all large $r$. Now since neither of Eq. \refer{eq1fsol1} and \refer{eq1fsol2} are satisfied for the entire $r$, according to Theorem \ref{nopiece} metric \refer{cbhmetric} in the SU(2) Yang-Mills-Einstein SSS model will allow the existence of timelike FP.

The second application of the theorems in this paper is that we can use them to study the existence and stability of the FPs of some general SSS metrics.
A total of 8 metrics from Ref. \cite{exactsolutionsofes}, whose $f(r)$'s are given in the first column of Table \ref{tab1}, are examined and their allowance of the timelike and null FPs are listed in the second and fourth columns respectively.
These spacetimes are chosen mainly because of their mathematical simplicity.
For metrics that do allow FPs, the columns three and five shows their stabilities obtained by using the sufficient conditions in section \ref{secstb} and \ref{secnull}, when they are applicable.
In column six we list  for the metrics allowing FPs whether they are asymptotically flat and if yes the sign of the corresponding Newtonian potential at large $r$. Apparently, as dictated by Theorem \ref{flatneg}, all asymptotically flat spacetime with negative effective Newtonian potential at large $r$ allows timelike FP. 
In Table \ref{tab1}, we also listed the most famous Schwarzschild and RN metrics, their FP existence and stabilities. These properties for these two metrics have been extensively studied elsewhere \cite{Pugliese:2010ps}
and it is seen that our results agrees very well with them. The difference in our approach is that after using the theorems in this paper, the calculations needed to obtain the same conclusion for these two spacetimes, and also for other metrics in Table \ref{tab1} which have never been studied before, become very elementary. This clears shows the power of our results.

\begin{table*}
\caption{The $f(r)$ component of the metrics, their existence of timelike FP (Yes:Y, No:N) and FP stabilities (Stable: S, unstable: U) are shown from column 1 to 3. The existence of null FP (Yes:Y, No:N) and FP stabilities (Stable: S, unstable: U) are shown in column 4 and 5. The asymptotic flatness of the metrics (Flat:F, non-flat: N) and sign of corresponding Newtonian potential at infinity (positive: +, negative: $-$) are shown in column 6. See Ref. \cite{exactsolutionsofes} for other components of the metrics except the RN metric whose $g(r)=f(r)^{-1}$. $a,~b,~n$ are real constants except otherwise stated. \label{tab1}}
{\small
\begin{tabular*}{\textwidth}{@{\extracolsep{\fill}}l||l|l||l|l||l@{}}
 \hline\hline
 $f(r)$&T. Exist.&T. Stab.&N. Exist.&N. Stab.&Asympt.\\
 \hline
   $1+a/r$          &Y $(a<0)$          &           &Y $(a<0)$          &U          &F, $-$     \\
                    &N $(a>0)$          &           &N $(a>0)$          &           &F, $+$     \\
   \hline
 $1+a/r+9b^2/(32r^2)$    &Y $(a<0)$          &           &Y $(a<-|b|)$ &        &F, $-$     \\
  ~~~$(b\neq0)$       &                   &           &N $(-|b|<a<0)$ &           &F, $-$     \\
                    &N $(a>0)$          &           &N $(a>0)$          &           &F, $+$     \\
   \hline
 $ar^b,~(a>0)$      &Y $(0<b<2)$        &S          &N $(0<b<2)$        &           &N          \\
                    &N $(b<0, b>2)$    &           &N $(b<0, b>2)$    &           &N          \\
   \hline
   $b^2+a^2r^2/b^2$ &Y                  &S          &N                  &           &N          \\
   \hline
$(a+br^n)^2,~(a>0)$ &Y $(b>0,n>1)$      &           &Y $(b>0,n>1)$      &S          &N          \\
                    &Y $(b>0,$          &S          &N $(b>0,$    &           &N          \\
                    &$~~~~0<n<1)$       &           &$~~~~0<n<1)$       &           &           \\
                    &Y $(b<0,n>0)$      &U $(n>1)$  &Y $(b<0,n>0)$      &S $(n>1)$  &N          \\
                    &N $(b>0, n<0)$     &           &N $(b>0, n<0)$     &           &F, $+$     \\
                    &Y $(b<0, n<0)$     &U $(n<-2)$ &Y $(b<0, n<0)$     &           &F, $-$     \\
   \hline
$a(5+br^2)^2(2-br^2)$& Y $(a>0)$        &           &Y $(a>0)$          &S          &N          \\
   ~~~$(ab<0)$      & N $(a<0)$         &           &N $(a<0)$          &           &N          \\
   \hline
   $r^2(a+b\ln r)^2$&Y                  &           &Y                  &S          &N          \\
   \hline
   $a\exp( br^2)$   &Y $(b>0)$          &S          &Y $(b>0)$          &S          &N          \\
   ~~~$(a>0)$       &N $(b<0)$          &           &N $(b<0)$          &           &N          \\
   \hline\hline
\end{tabular*}
}
\end{table*}

Regarding ways to extend the current work, there exist at least a few. The first is to notice that Eqs. \refer{eqone} and \refer{eqtwo} and all the analysis followed do not rely on the dimension of the spacetime \cite{Tangherlini:1963bw}. Therefore all conclusions given by the Theorems in this paper also apply to SSS spacetime of higher dimension. Since these spacetimes and the geodesics in them are of their own importance \cite{Hackmann:2008tu}, our work can shed light on the existence of COs and their stabilities in these spacetime. The second way is to use our result to study the COs of some other important SSS spacetime, especially those metrics whose closed form is not yet obtained. From the SU(2) Yang-Mills-Einstein example we gave above, and from the Eqs. \refer{eq1fsol1} and \refer{eq1fsol2}, it is clear that the existence of COs does not require the complete knowledge of the analytical metric functions on all $r$, but their dependence on $r$ and their asymptotics should be enough. Since enormous difficulties usually lie in the solution process from these dependence to the analytical form of the metric, our results can be an important tools for determine the existence and stabilities of COs for these metrics.
The last extension is to relax the SSS requirement but to study using a similar strategy the existence and stability of FPs in equatorial plane of axially symmetric spacetime. Since a general form of an axially symmetric metric, which is the only starting point that our analysis needs, is already known in the Weyl-Lewis-Papapetrou coordinates \cite{Weyl:1917gp, dewitt}, we expect that an analogous approach should also be applicable here. Research in these directions are now pursued.

\ack
We appreciate the discussion with Dr. Hongbao Zhang.  The work of J. Jia, J. Liu, X. Liu and X. Pang are supported by the Chinese SRFDP 20130141120079, NNSF China 11504276 \& 11547310, MST China 2014GB109004 and NSF Hubei ZRY2014000988 /2014CFB695. The work of N. Yang is supported by the NNSF China 31401649 \& 31571797.

\end{document}